\documentclass[conference]{IEEEtran}

\usepackage[acronym]{glossaries}

\usepackage{diagbox}
\usepackage{cite}
\usepackage{array}
\usepackage{tabularx}
\usepackage[table]{xcolor}
\usepackage{pgfplotstable}
\usepackage{smartdiagram}
\usepackage{mathtools}
\usepackage{amssymb, amsmath}
\usepackage[hyphens]{url}
\usepackage{epstopdf}
\usepackage{algorithm}
\usepackage{algpseudocode}
\usepackage{expl3}
\usepackage{url}
\usepackage{lipsum}

\usepackage{soul,color}
\soulregister\cite7
\soulregister\ref7
\soulregister\pageref7

\makeglossaries
\ifCLASSOPTIONcompsoc
\usepackage[caption=false,font=normalsize,labelfon
t=sf,textfont=sf]{subfig}
\else
\usepackage[caption=false,font=footnotesize]{subfi
	g}
\fi
\ifCLASSINFOpdf

\else

\fi

\begin{document}

\title{A wrinkle in time: A case study in DNS poisoning}
\author{Harel Berger \and Amit Z. Dvir \and Moti Geva}

\maketitle 

\begin{abstract}
The Domain Name System (DNS) provides a translation between readable domain names and IP addresses. The DNS is a key infrastructure component of the Internet and a prime target for a variety of attacks. One of the most significant threat to the DNS's wellbeing is a DNS poisoning attack, in which the DNS responses are maliciously replaced, or poisoned, by an attacker. To identify this kind of attack, we start by an analysis of different kinds of response times. We present an analysis of typical and atypical response times, while differentiating between the different levels of DNS servers' response times, from root servers down to internal caching servers. We successfully identify empirical DNS poisoning attacks based on a novel method for DNS response timing analysis. We then present a system we developed to validate our technique that does not require any changes to the DNS protocol or any existing network equipment. Our validation system tested data from different architectures including LAN and cloud environments and real data from an Internet Service Provider (ISP). Our method and system differ from most other DNS poisoning detection methods and achieved high detection rates exceeding 99\%. These findings suggest that when used in conjunction with other methods, they can considerably enhance the accuracy of these methods.
\end{abstract}
\section{Introduction}
\begin{figure}[!h]
	\centering
	\includegraphics[width=0.7\columnwidth]{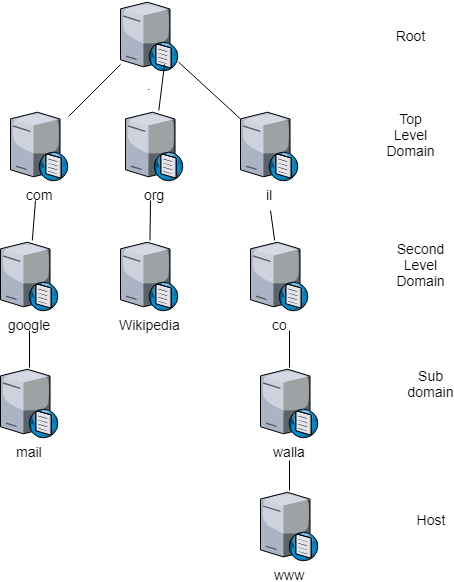}
	\caption{DNS hierarchy example}
	\label{fig:hierarchy}
\end{figure}

The Domain Name System (DNS) \cite{rfc1034,rfc1035} is one of the best known protocols in the Internet.  Its main function is to translate human-readable domain names into their corresponding IP addresses. Its importance for the Internet is derived from the fact that virtually all day-to-day network applications use DNS.
The translation process is done through DNS queries between the client and the DNS server (resolver). The ordering of the DNS tree, from the root down, is called the "DNS Hierarchy" and is depicted in Fig. \ref{fig:hierarchy}. 

There are two types of DNS resolvers, termed Authoritative and Recursive. Authoritative name servers give answers in response to queries about IP addresses. They only respond to queries about domains that have been configured to provide answers. Recursive resolvers provide the proper IP address requested by the client. They do the translation process by themselves, and return the final response to the client. In this paper, we focus on recursive DNS resolvers. 

Generally, clients issue their queries using DNS messages to a DNS resolver which maps each query to a matching Resource Record (RR) set and returns it in the response DNS message. Each record is associated with a Time-To-Live (TTL) value. Resolvers are allowed to store (cache) the response in their memory until the TTL value expires. When this time period has elapsed, the RR is evicted from
the cache. 

Given a query to resolve, a recursive resolver looks up the cache for a matching record. If one exists, it is returned as the response. If not, the recursor uses the DNS resolution process to obtain a matching record by implementing the following steps. First, it determines the closest zone (level in the hierarchy) that encloses the query and has its information cached. If no such zone is cached, the enclosing zone is the root zone. In this case, the recursive resolver resorts to contacting the DNS root-servers. The root server it contacts returns an authoritative response, which redirects the recursive resolver to a Top Level Domain (TLD) server. Then, the recursor requests a response from the TLD server. The TLD server sends the full resolution response, or a redirection to a Second Level Domain (SLD) server. A process similar to the one for the TLD is carried out by the SLD server, and the DNS hierarchy levels beneath it (in the DNS hierarchy tree).

Various studies have examined DNS attacks\cite{bt2017,kambourakis2008detecting,jackson2009protecting,cheshire2013dns,ballani2008mitigating,son2010hitchhiker,banu2013comprehensive} designed to prevent clients from resolving RRs. One of the most common attacks is DNS cache poisoning, where the attacker provides spoofed records in the responses to redirect the victims to malicious hosts. This kind of attack can facilitate credentials theft, malware distribution, censorship and others.

Over the years, a number of improvements have been suggested to secure the DNS. One of the main advances is known as Domain Name System Security Extensions (DNSSEC) \cite{rfc4033,rfc4034,rfc4035}. It consists of a set of extensions to the DNS which provides DNS resolvers with origin authentication of DNS data, authenticated denial of existence, and data integrity. All responses that use DNSSEC are digitally signed. By checking the digital signature, a DNS resolver is able to verify whether the information is identical (i.e. unmodified and complete) to the information published by the zone's owner and served on an authoritative DNS server. In so doing, DNSSEC protects against cache poisoning \cite{silva2009dnssec}. 

DNSSEC appears to be efficient, but it does not guarantee availability or confidentiality.
Its deployment rate is less than 20\% of all name servers globally, as updated monthly in the Internet society's statistics\cite{stats_lab}. In addition, many DNSSEC keys are vulnerable \cite{dai2016dnssec}. Thus shielding from cache poisoning remains a crucial unresolved issue, as described in \cite{anu2017survey}. 

\label{Intro}

\begin{figure}[!h]
	\centering
	\includegraphics[width=1\columnwidth]{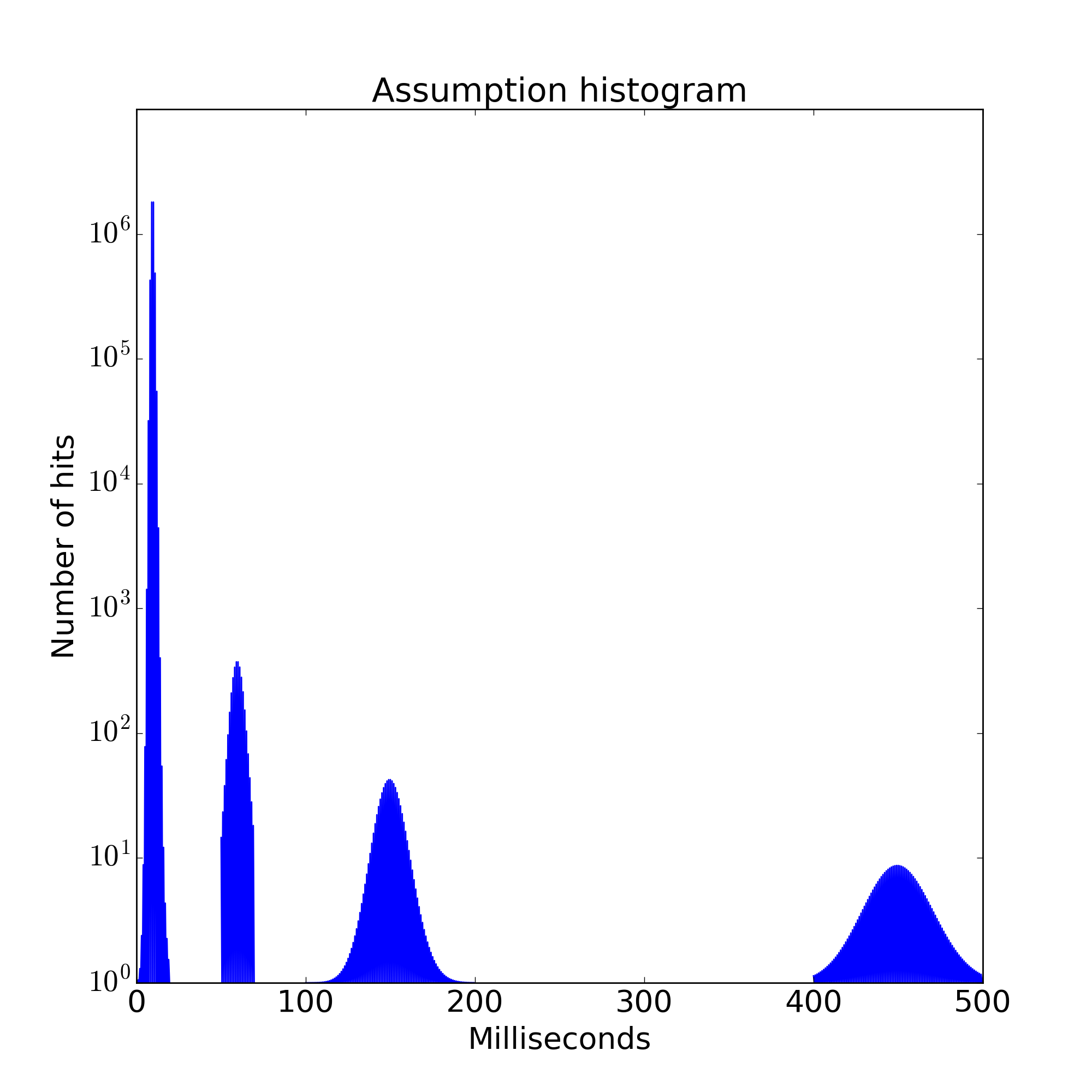}
	\caption{We hypothesized that the distribution of the DNS time value would consist of a series of distributions. To simplify the presentation, we assume they are a union of Poisson distributions. Between each two Poissons there is a gap with no values. Therefore an attack that generates a time value located in one of the gaps or proximal to it can be pinpointed with high probability.}
	\label{fig:assume}
\end{figure}

\noindent\textbf{Contribution of This Work:} Our hypothesis, as shown in Fig. \ref{fig:assume}, is that the time elapsed from the DNS query to the relevant DNS response is best fit by a group of Poisson distributions. A Poisson distribution is a statistical distribution of the likelihood an event will occur within a time interval. The average number of events in an interval is designated by $\lambda$. A Poisson distribution is a typical decision in the field of network attacks\cite{mann2011reactive}. In this paper, each Poisson resembles a level in the DNS resolution process. We posited that the gaps between the Poissons could indicate  anomalies. Thus anomalies point to attacks with high probability. 

To test this assumption, we chose a sample of Alexa's top sites\cite{alexa}. We acquired data from simulations and from a real ISP. We inspected the distribution of the entire domains' data and each domain separately. Some of the cases we examined exhibited a distribution similar to our assumption. We created attack data, based on a third party attack tool. We constructed a detection system for the attacks based on the simulation data. 

Our key contribution is the innovative notion that attack detection can be based solely on time value. This simple concept is also efficient in terms of running time and memory. Our simulations ran at different times, and on different network parameters, environments etc. Therefore, our data includes jitter and packet loss. In our analysis we do not mention this noise explicitly. Furthermore, the levels detection in Section \ref{spec} is compared to a real data from an ISP.
The other novel insights we present here are as follows:
\begin{itemize}
\item Timing analysis of DNS authoritative servers - We  measured the response times of the root and TLD levels. We executed this from  two vantage points, one of which was local(Ariel University, Israel) and the other from Google cloud(St. Ghislain, Belgium). Our method of time measurement is configuration dependent, but can easily be generalized to other cases. We analyzed each server separately to define its time value distributions. Although each server had its own time distribution, the Poisson distribution was identified in each of them.
We report our analysis of both the local and the cloud environments alongside real data from Israeli universities' ISP\cite{Inter}.

\item The development of a detection system for DNS poisoning attacks - We describe a detection system based on observations and their interpretation. The system does not require any changes to the protocol or any existing network equipment. It can be used as a standalone in any network to obtain a classification of DNS responses, and better ensure the identification/prevention of DNS attacks.

\item Testing the accuracy of the system based on simulation and empirical analyses - We tested an actual cache poisoning tool built by a third party on our system. Our accuracy rate exceeded 98\%.
\item Mitigation of success rate - We compared the success rate of the attack with and without our system. We were able to reduce the attacker's success rate by 70\%-80\%.
\end{itemize}

The remainder of this paper is organized as follows.
We discuss previous works on DNS measurement and cache poisoning attacks in Section \ref{relate}. Then, we present our model in Section \ref{model}. In Section \ref{Meth} we describe how the method was implemented to analyze DNS packets in a local network and in the cloud. We differentiate between cache and non-cache responses and use them to show ways to differentiate between DNS levels with high accuracy. In Section \ref{Ident} we describe a novel poisoning attack identification method.
We discuss several learning machines we used for the methodology and the identification of the attack in Section \ref{Learning}. The technical specifications appear in Section \ref{Expo} and the results in Section \ref{results}. Section \ref{conclu} concludes this paper.
\section{Related Works}
\label{relate}
In this section, we summarize previous studies related to DNS measurement and cache poisoning attack identification.

Van Rijswijk-Deij et al.\cite{van2016high} surveyed a large variety of TLD servers. Their overview spanned features of DNS measurements such as duration, goals, number of vantage points, etc. However, their tests cases only examined cloud email services. 

Ager et al.\cite{ager2010comparing} evaluated the response time of ISP DNS resolvers, GoogleDNS\cite{Google_dns} and openDNS\cite{Open_dns}. 
This study analyzed the time value of DNS resolvers, but disregarded DNS hierarchy levels. None of these works which analyzed DNS server measurements, considered measuring the DNS levels from the root down to the internal caching resolvers or  the distribution of DNS time values. 

Wang et al. \cite{wang2006tracking} proposed an associative feature analysis approach based on statistical models to track the anomalous behavior of DNS servers. In collaboration with a major commercial ISP in China, they captured and analyzed real DNS traffic in this large-scale network environment. They used an outlier function to map  malicious responses. The parameters they used were queries/responses per client/server/specific server. The authors detected various attacks in the real world, but did not determine the real volume of the attacks. In other words, they could not evaluate their accuracy.

Yamada et al. \cite{yamada2009anomaly} focused on an anomaly detection system for DNS servers. Normally, dealing with large number of hosts can consume vast amounts of computational resources and make real-time analysis difficult due to traffic overload. They proposed anomaly detection for DNS servers that frequently invoke host selection in which only potential hosts are selected.
They used a FIFO (First In First Out) based method for frequent host selection along with other statistics. They categorized packets by type such as DNS mail records (MX), regular DNS packets (A), error rate, etc. They proposed several heuristics, such as the number of queries and requests, where a slightly higher rate of queries/requests was considered to indicate an attack.
They identified attacks such as the spam Backscatter. This kind of spam consists of incorrectly automated bounce messages sent by mail servers, typically as a side effect of incoming spam (unsolicited messages). Their attack identification system achieved 68\% accuracy.

Klein et al. \cite{kl2017} focused on cache poisoning attacks. They investigated a new  indirect attack where they injected the victim's cache with a poisonous record which does not immediately impact the resolution process, but rather becomes effective after an authentic record expires. In this case, the next resolution request to that name  returns the spoofed record. Canonical NAME record (CNAME) is a type of resource record in the DNS used to specify that a domain name is an alias for another domain called the "canonical" domain. Delegation of the NAME record (DNAME) creates an alias for an entire subtree of the domain name tree. They injected CNAME and DNAME responses in a cache poisoning attack.

Celik and Oktug \cite{celik2013detection} researched the Fast-Flux Service Networks (FFSN) method used by bots. An Internet bot is a software application that runs automated tasks (scripts) over the Internet. Typically, bots perform tasks that are both simple and structurally repetitive at a much higher rate than would be possible for a human alone. Fast flux is a DNS method used by bots to conceal their actions behind an ever-changing network of proxies. Bots use FFSN to hide phishing and malwares through networks of proxies and servers.
The authors used a variety of features for machine learning and achieved 98\% accuracy in FFSN identification. Some of the features were taken from the DNS packets themselves, such as the number of unique A records and NS records. They also mapped the A and NS records to their geographical identifiers to better understand their spatial entropy. They counted the AS related to each IP and inspected the RTT values, but failed to generate a proper analysis because of the processing and delays they could not dissect.
 
An Intrusion Detection system (IDS) which is not specific to DNS but is related to our work was implemented in
Ertoz et al. \cite{ertoz2004minds}. They used outliers as identifiers of anomalies to detect attacks such as port scans, worms, etc. (no list is provided in the article). They also explored the non-authorized use of protocols (without inspecting the payload).
The features were divided into three parts: the package header, the time window statistics, and the connection statistics.
The anomaly detection module was the Local Outlier Factor (LOF). The outlier factor of a data point was local in the sense that it measured the extent of being an outlier with respect to its nearest data points. Each new data point distance was compared to the density of the class data points. If its distance was smaller, it was not considered an outlier.
A pattern matching method was also used on the top 10\% suspicious connections in the previous method to identify future attacks.

Overall, these studies used statistical methods such as outliers and LOF to identify attacks. Some used pattern analysis based on prior knowledge. However, none implemented a time stamp/time analysis as the main detection strategy. Nevertheless, time based analysis is easy to use, and efficient in terms of memory and number of calculations. Recording RTT values (along with domain names) is simple to do, which makes this identification method easy to deploy in a large number of systems.   
\section{The Model}
\label{model}
\begin{figure}[htp]
\centering  \includegraphics[clip,width=1\columnwidth]{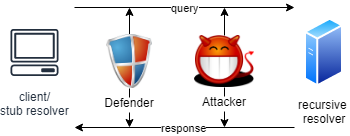}%

\caption{Our model is composed of the client, recursor, attacker and defender. The defender and attacker are located inside the LAN.}
\label{fig:model}
\end{figure}

Our model consists of a client, a recursive resolver, an attacker and a defender. The defender possesses an additional offline dataset of DNS benign responses. Fig. \ref{fig:model} depicts this model. 
The attacker is described in Section \ref{at_mod}, and the defender in \ref{def_mod}.

\subsection{Attacker Model}
\label{at_mod}
Our attacker module is an eavesdropper that can also inject malicious packets \cite{yao2014network}. It cannot drop a packet it sees. It can respond with a fake packet to a query it sees. In addition, the attacker cannot access the offline data of the defender.  

Typically, the attacker in DNS poisoning attacks is an off path attacker\cite{herzberg2013socket,herzberg2013dnssec,eastlake2000rfc} which is considered to be a more realistic attack model. Getting access to a LAN or attacking on-path between a recursive resolver and authoritative name server is much harder than spoofing response packets. The success rate of an attacker that spoofs packets is low since it needs to guess certain random parameters, such as the query ID and port number. However, it can acquire these with an additional puppet\cite{herzberg2013socket}, or through flaws in the resolver software design\cite{klein2007bind}, etc. BGP Hijack attacks \cite{al2016bgp,huston2010securing} enable the interception of traffic, including DNS queries. In this frequent type of attack, the attacker who has this ability can counterfeit DNS responses effortlessly. 

Our attacker is placed inside the LAN. This defines our attack as DNS poisoning against the client, rather than the notorious cache poisoning attack. However, our identification method does not need modification to enhance its ability to detect a cache poisoning attack. The only difference is where the attacker and defender are placed.

\textbf{The attacker's success} is defined as the case where the attacker's response is categorized by the defender as a benign response. \\
\textbf{The attacker's failure} consists of sending a counterfeit response that the defender correctly labels as an anomaly/malicious.
\subsection{Defender model}
\label{def_mod}
Our defender's module is a sniffer. It classifies any new sample as a benign or a malicious response. It has an offline dataset of domain names and RTT values. It classifies the new samples according to its dataset. 
We assume that it sniffs the domain names of each response and other network parameters such as the RTT value. We use the time value as the main feature of this study to map the resolve levels and their time differences. In so doing we reduce analysis time. 

\textbf{The defender's success} is defined as the case where the attacker's response is labeled as a malicious packet.\\ 
\textbf{The defender's failure}  is the case where the attacker's response is erroneously labeled as a benign packet.
\section{Methodology}
\label{Meth}
This section presents the data analysis methodology. We had two sources for the data in this study. The first was the experiments we conducted. We used local and cloud environments. Our experimental data are fully described in Section \ref{Expo}. We aimed to confirm our hypothesis not only on the simulation data but also on real data. Therefore, we acquired a second source data from the Inter-University Computation Center (IUCC) .  The IUCC serves a vast number of faculty members and students at Israeli universities and regional colleges as well as researchers in numerous R\&D organizations in Israel. IUCC is considered Israeli universities' ISP. 




The data for our experiments were recorded from both a stub resolver and a recursive resolver. We mapped each query from the client to the correlative query that arrives at the recursive resolver by the ID and domain parameters. A combination of these two parameters produced a one-to-one mapping of the client's queries and the recursor's responses. We did the same for the fully resolved response from the recursive resolver to the client. Therefore, we could have inspected the resolution process as a whole. However, the data from the IUCC were recorded from one link between the recursive resolver and the authoritative name server. Some of the queries got responses in another link. Some responses we saw were for queries that were sent in an alternative link. Hence we could not aggregate the resolution process.  Also, we did not have any control over which queries were asked during the record session, which domains were asked etc. Therefore, we could not correlate the domains from the IUCC to the domains from our experiments. As a result, these data are mainly discussed at the Specific Level Analysis (Section \ref{Spec}). 

The first step was to separate the cache and the resolve responses in the data to identify which response was responded to by the recursive resolver's memory and which ones were either given a full or partial resolution process. This step is important for two reasons. First, because our methodology tends to maps time values to resolution processes, we map the lack-of-resolution time interval. Second, separating it from the fastest resolution process indicates a time interval where no responses arrive. An attacker's packet that arrives during this time interval can thus be distinguished easily from cache or resolved responses.

\begin{figure}[!h]
	\centering
	\includegraphics[width=1.01\columnwidth]{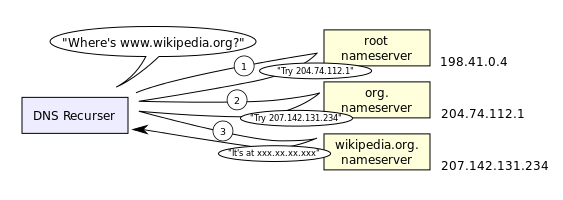}
	\caption{DNS resolve levels for www.wikipedia.org. The resolving process is done by the DNS recurser (recursive resolver) on www.wikipedia.org. The resolution takes place in steps 1-3 by the root, org. and wikipedia.org. nameservers.}
	\label{fig:rec_levels}
\end{figure}

After separating the cache and resolve levels, we broke down the DNS resolve levels. The DNS level of a query sent to the recursive resolver is the highest level in the DNS hierarchy to which any of the resolution queries were sent by the resolver. For example, if a query about www.wikipedia.org was sent to the recursive resolver (as in Fig. \ref{fig:rec_levels}), and it sent a resolution query to the root server (which is the highest level in the hierarchy), the  query was tagged as root level. 

Then, we analyzed each DNS level separately to better understand the RTT distribution. We inspected each domain by separating the data into domains. This provided a clearer view of the distribution of each domain. Each domain was inspected in terms of the DNS levels, and yielded a probability table for each domain and the specific time intervals in each.

We analyzed the data in four steps:
\subsubsection{Cache or Resolve analysis} This analysis identified which response was from the cache and which one had received a full/partial resolution process. 
\subsubsection{Hierarchy Levels Analysis} This analysis separated the DNS resolve levels over the entire dataset.
\subsubsection{Specific Level Analysis} This analysis focused on the distribution of each DNS level.
\subsubsection{Specific Domain Inspection} This analysis grouped the data into domains to get a clearer view.

As mentioned in Section \ref{model}, our attacker attacks a client. This analysis allowed us to determine from the client's point of view how the data distributed for each DNS level. From the client's point of view, every response is received from the same IP - the recursive resolver's IP. Therefore, we need to distinguish between DNS levels. Note that in a cache poisoning attack, there is no need for our methodology. In this case, the data for each DNS level are separate, since the recursive resolver sees the IPs and can track each DNS level separately. Therefore, this section is not relevant for a cache poisoning attack. However, Section \ref{Ident} can be used both for our attack and for cache poisoning attack.

\subsection{Cache or Resolve analysis}
\label{Cache_or_non}
\begin{figure}[!h]
	\centering
	\includegraphics[width=1.01\columnwidth]{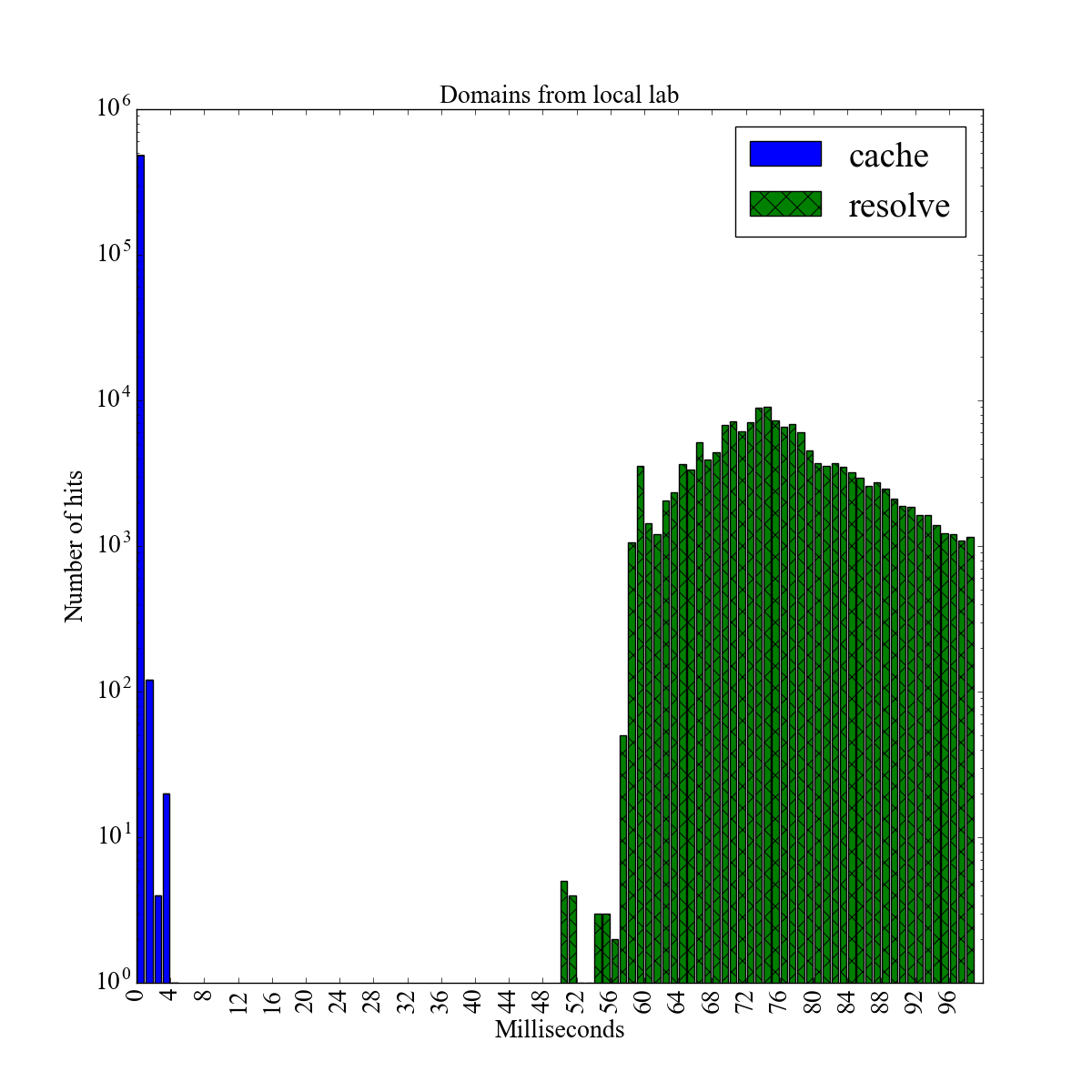}
	\caption{Classification of recursive resolver responses, between 0-100 ms, at 1 ms intervals. The estimated ping value is 2 ms, because the recursive resolver in this experiment was in the same LAN of the client. Most of the smooth responses (cache) were lower than the ping value. All the hatched circle responses were higher.}
	\label{fig:cache-non}
\end{figure}

First, the cached responses were separated from the resolved responses. We assumed this would be feasible since the resolve process takes time.
We found that there was a considerable difference between the cached responses and the resolved responses as a function of the components and the communication system. Fig. \ref{fig:cache-non} depicts these differences, where the cached responses are on the far left side and the resolved responses are on the right side. A gap of $\sim$50 ms appears between them.

The cached responses were identified by the fact that the query from the client and the response from the recursive resolver were successive. As a further confirmation method, we tested the ping between the client and the recursive resolver, and compared the average value of the ping to the actual response values. The response RTT values that were close to the average ping value were from the cache.

Fig. \ref{fig:cache-non} depicts the distribution of the RTT values between 0-96 ms. acquired from one of our experiments. The figure shows that there was a wide gap between the cache on the left side (blue) of the histogram, and the beginning of resolve responses on the right side (green) of the histogram.

This analysis led us to the conclusion that the cached responses could be easily identified. The next step was to determine whether we could identify each level of the DNS hierarchy.

\subsection{Hierarchy Levels Analysis}  
We inspected the behavior of the DNS levels.
Each resolution query from the recursive resolver was inspected individually, which yielded detailed data about the client's query it was connected to. By identifying the source's IP from the response, it was possible to identify the current level of resolve of the resolve query. Each response's source IP was identified in a reverse DNS query to get its DNS level. We mapped each client's query to the highest DNS level of any resolution query that was done for it. It was impossible to extract the DNS levels' distributions from the data as a whole since most of the intervals were mixed with a number of DNS levels. We had to devise another way to approach  the problem.


As a result, our next step was to separate the levels and inspect them individually. In each level, we took the relevant authoritative servers and verified the RTT values based on the responses acquired from them.
\subsection{Specific Level analysis}
\label{spec}
For this analysis, we examined both our data sources. We separated out each resolve query to map each DNS level. We generated histograms for each authoritative server at the specific level included in our data to fully examine their RTT distribution. These histograms can vary from different vantage points. To test whether our hypothesis on the RTT distribution would hold with respect to each vantage point, we ran our experiment from two vantage points. The third vantage point was the IUCC. 

\begin{figure*}[!]
    \subfloat[Root servers distributions from the local experiment.]{\includegraphics[width=0.66\columnwidth]{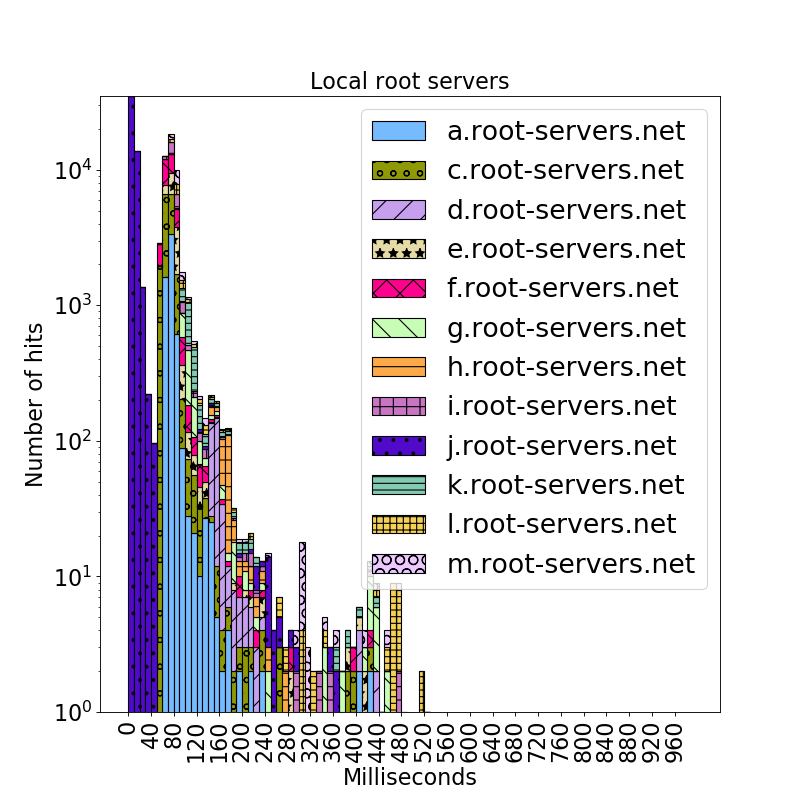}}
		\label{roots_local}
		\hspace{0.5cm}
	\subfloat[Root servers distributions from the cloud experiment.]{\includegraphics[width=0.66\columnwidth]{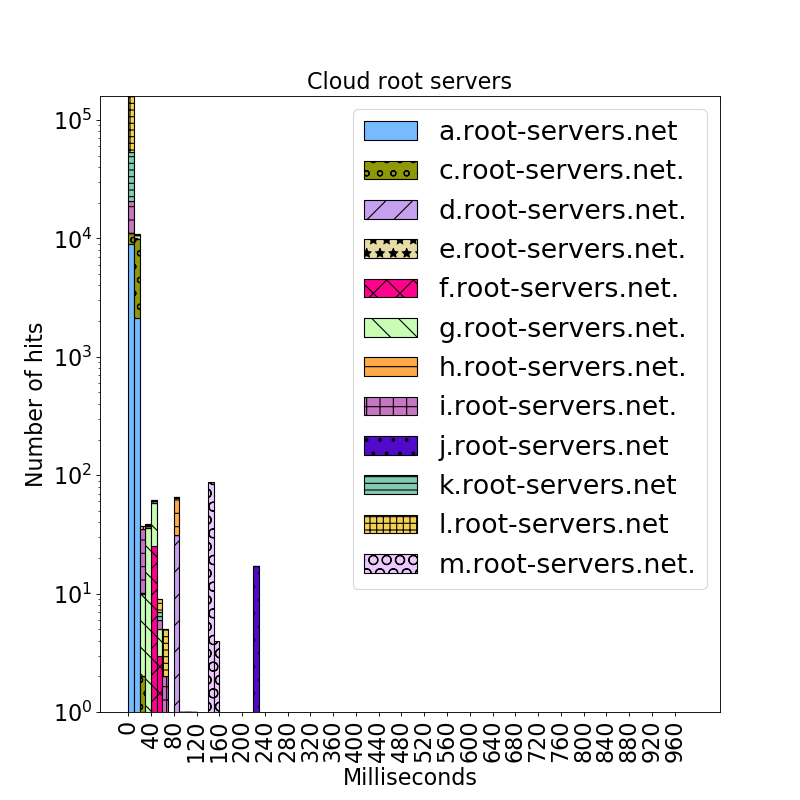}}
		\label{roots_go}
		\hspace{0.5cm}
	\subfloat[Root servers distributions from the IUCC data source.]{\includegraphics[width=0.66\columnwidth]{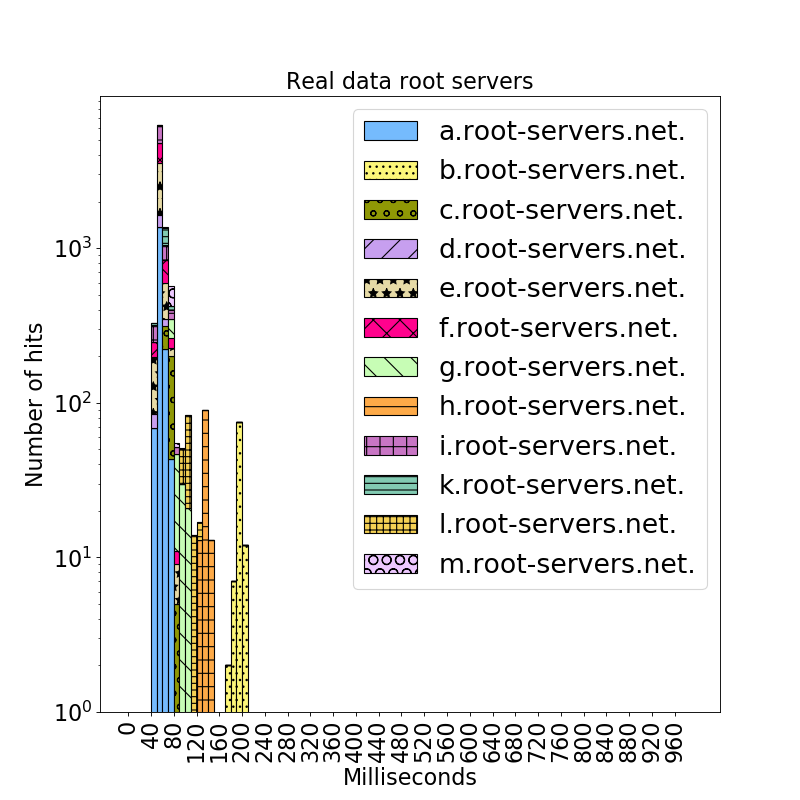}}
		\label{roots_mach}
	\subfloat[gTLD servers distributions from the local experiment.]{\includegraphics[width=0.66\columnwidth]{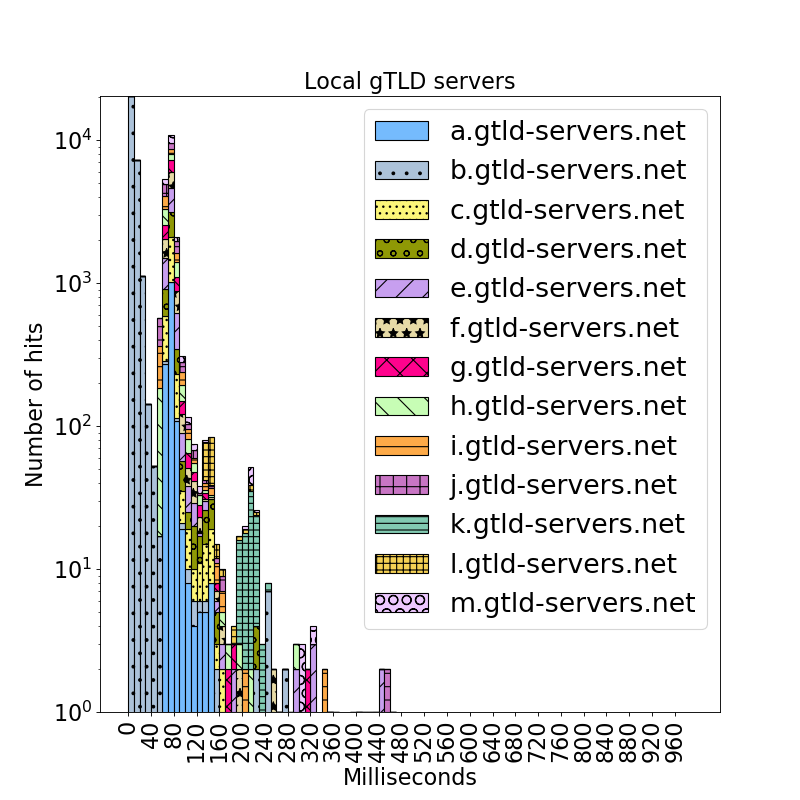}}
		\label{gtlds_local}
		\hspace{0.5cm}
	\subfloat[gTLD servers distributions from the cloud experiment.]{\includegraphics[width=0.66\columnwidth]{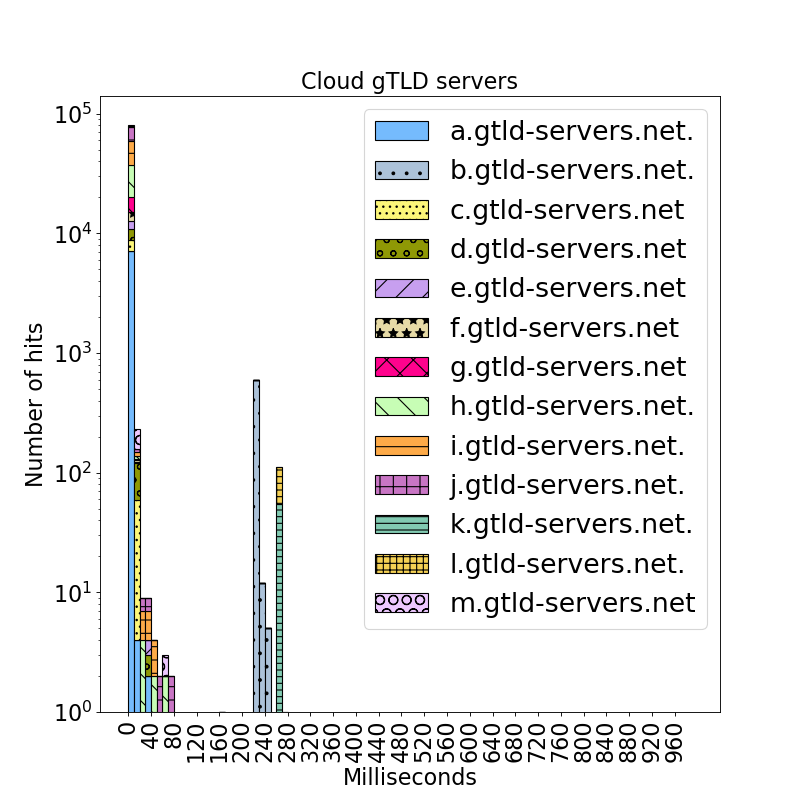}}
		\label{gtlds_go}
		\hspace{0.5cm}
	\subfloat[gTLD servers distributions from the IUCC data source.]{\includegraphics[width=0.66\columnwidth]{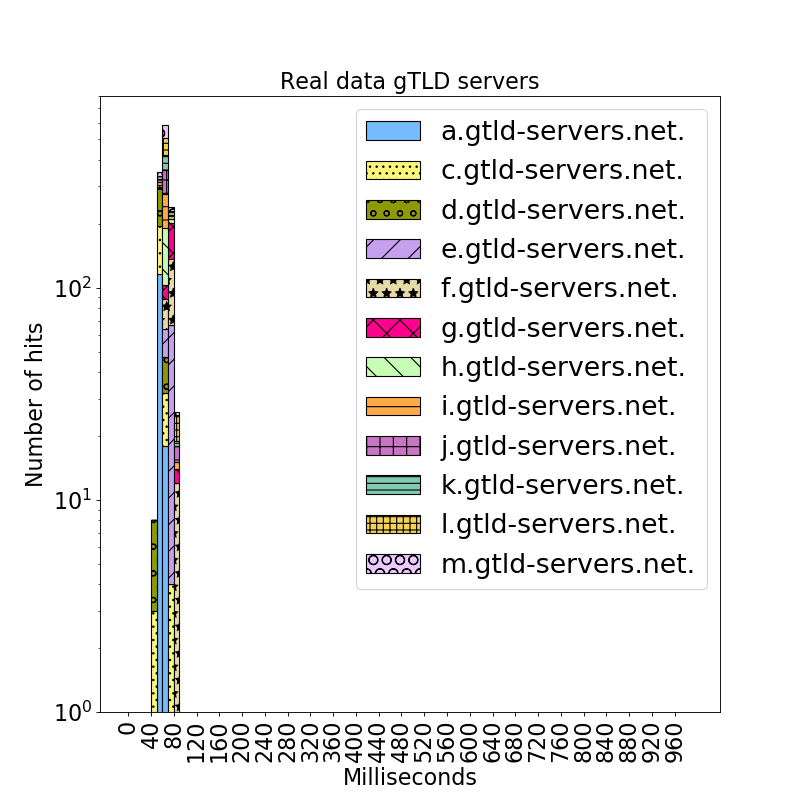}}
		\label{gtlds_mach}
    \subfloat[ccTLD servers distributions from the local experiment.]{\includegraphics[width=0.66\columnwidth]{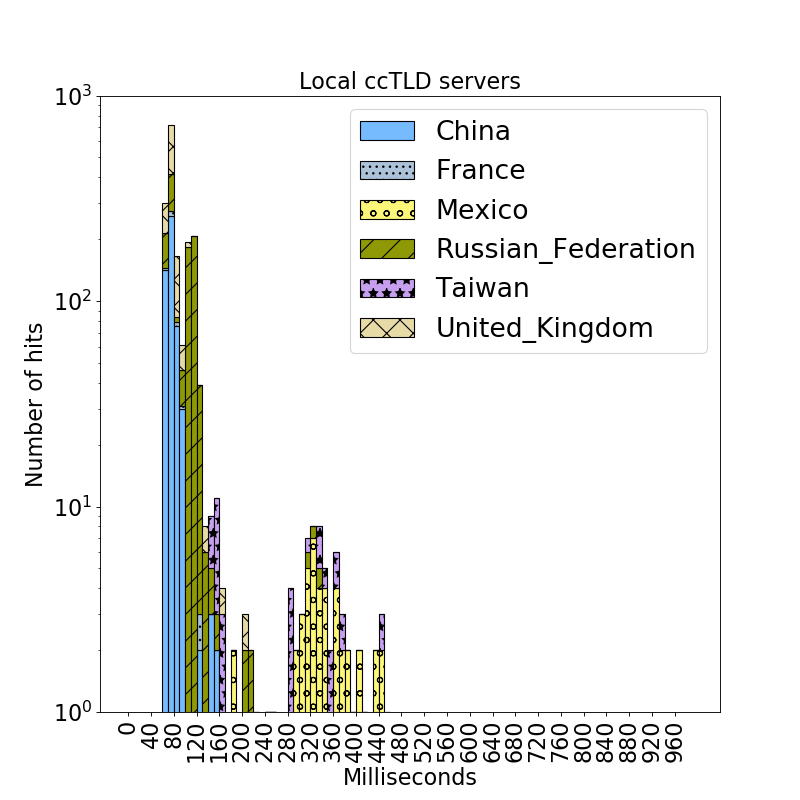}}
		\label{cctlds_local}
		\hspace{0.5cm}
	\subfloat[ccTLD servers distributions from the cloud experiment.]{\includegraphics[width=0.66\columnwidth]{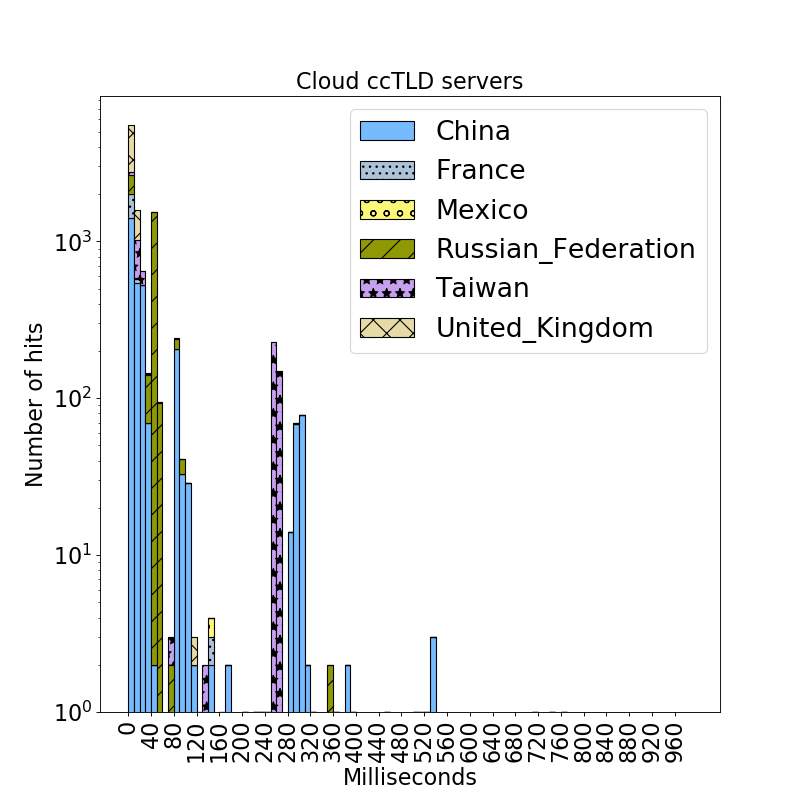}}
		\label{cctlds_go}
		\hspace{0.5cm}
	\subfloat[ccTLD servers distributions from the IUCC data source.]{\includegraphics[width=0.66\columnwidth]{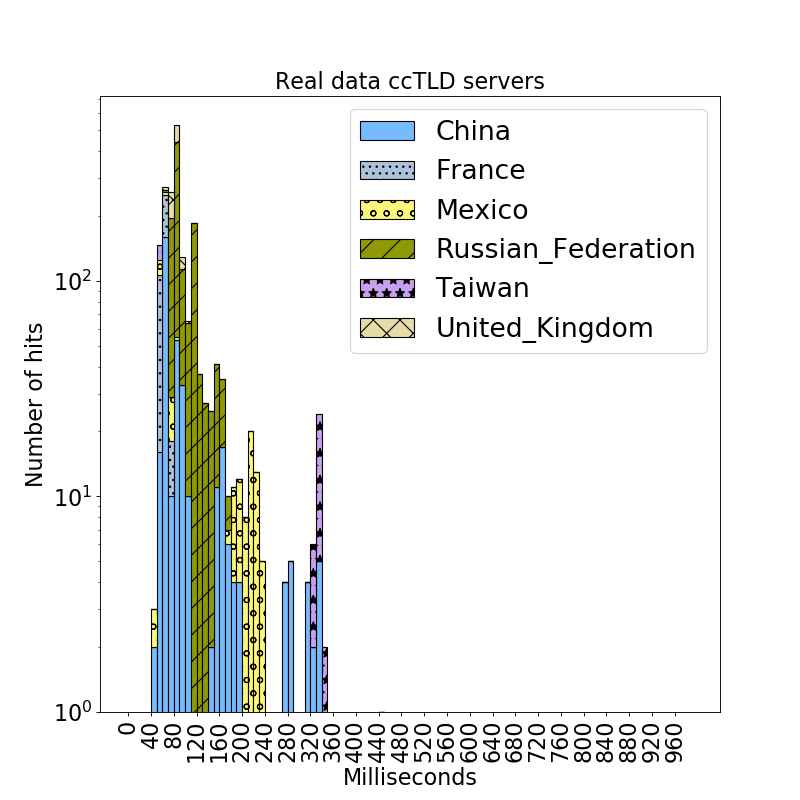}}
		\label{cctlds_mach}
		
\caption{Local and cloud experiments had no any data from root server b. IUCC did not include any data from root server j or gTLD server b. The majority of RTT values from the root servers were between 80-160 ms in the local experiment. In the cloud experiment, the majority were between 0-40 ms. Most of the responses from root servers arrived between 40-80 ms in the IUCC data source. As for the gTLD servers, the majority of the RTT values in the local data experiment were between 80-120 ms. (except from server b). In the cloud data experiment, the majority was between 0-40 ms. Most of the responses from gTLD servers arrived between 40-80 ms in the IUCC data source. We picked 6 countries for the ccTLD servers: China, France, Mexico, Russia, Taiwan, and the UK. Each country had a unique distribution. China was the most heterogeneous.}

	\label{fig_servers}
\end{figure*}

First, we looked at the behavior of the root level. We created histograms for the 13 root servers\cite{iana_roots}, as depicted in Fig. \ref{fig_servers}(a-c). Except for root server j in the local  experiment, most of the responses arrived within an interval of 40-80 ms, with differences of 40 ms between them. Root server j was inspected separately and was found to be much closer to the recursive resolver than the other root servers. The average ping value of root server j was $\sim12$ ms, whereas the minimal ping value of other root servers was $\sim80$ms. In addition, a traceroute check showed that it was the only root server whose route included IIX ISP, which was located near our local experiment site. These features confirmed that root server j was the most efficient root server for the local experiment.

Second, we examined the behavior of the TLD level to determine the differences between the behavior of the general TLD (gTLD) level servers, and the country code TLD (ccTLD). The country code TLDs are reserved for a specific country or state. We separated the country code TLDs from the other TLDs to determine whether there was a difference in their response times. We generated histograms for both types of TLD servers as shown in Fig. \ref{fig_servers}(d-i).

Because there were too many gTLD servers to depict in one histogram, we chose one of the most frequently used gTLD server types (from a.gtld-servers.net to m.gtld-servers.net) that is representative of .com and .net domains. As depicted in Fig. \ref{fig_servers}, the histograms were similar to the root servers, especially in terms of the mean and the distribution of the RTT values. Similar to the root servers, there was one server in the local data experiment that was different. The b.gtld-servers.net ping value was $\sim15$ms, whereas the other gTLD server's ping value was at least $\sim80$ms. Furthermore, the traceroute command showed a similar result as for root server j.

Country code TLDs are scattered over the globe. Therefore, their RTT values depends on the country and the distance from the recursive resolver. We took 6 countries as an example, and created three histograms for every vantage point. The histograms are depicted in Fig. \ref{fig_servers}(g-i). Clearly, the ccTLDs RTT values were different across countries and the distribution within each nation was different. Most of China's RTT values were located between 50-90 in the local experiment and IUCC, whereas Russia's was 120-160 ms in the local experiment, and 80-120 ms. in IUCC. 

The results showed that every level had its own RTT distribution. The synthetic data we obtained from our experiments showed similarities to the actual data from the IUCC in Fig. \ref{fig_servers}(c) and \ref{fig_servers}(f), with small shifts between them. Thus, the distribution of the synthetic data proved to be equivalent to the real data. The distributions of the root and the gtld servers were similar to a Poisson distribution, whereas the ccTLDs were diverse (due to geographical distances). 

We next tested our predictions on domains, to make the attack identification more precise. By separating into domains we expected to find more gaps in the RTT distribution. We also tested each part of the Poisson distribution. Thus, in the next stage we separated the data into domains.
\subsection{Specific domain inspection}
\label{Spec}
The data in this inspection were separated by domain. We inspected the top 500 domains from alexa.com. Each domain was inspected individually, resulting in $\sim500$ smaller distributions. 

\begin{figure*}[!th]
	\centering
	\subfloat[Youtube.com distribution]{\includegraphics[width=2.5in]{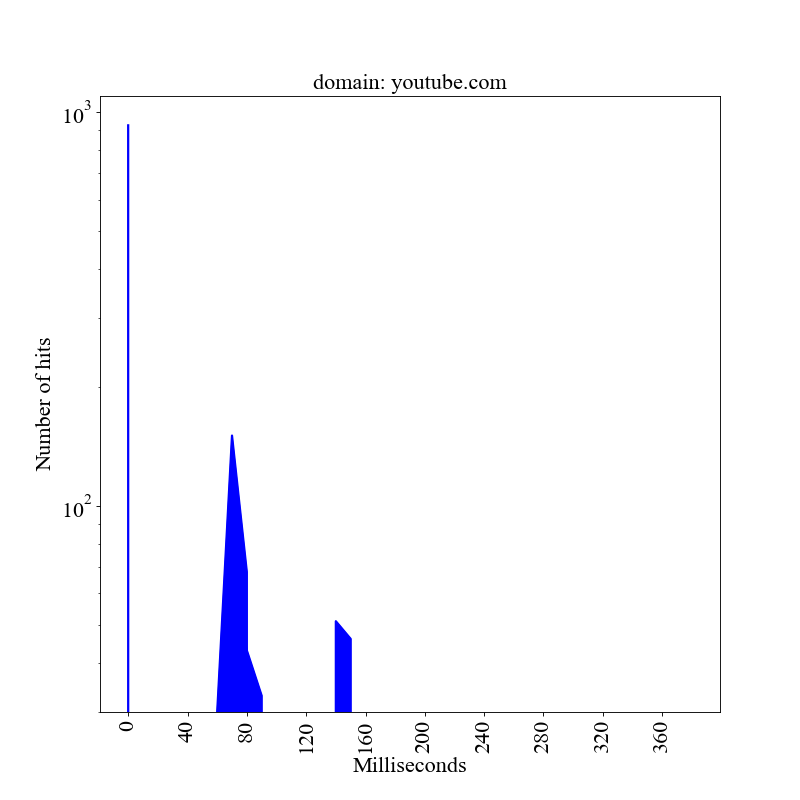}}
    \subfloat[Facebook.com distribution]{\includegraphics[width=2.5in]{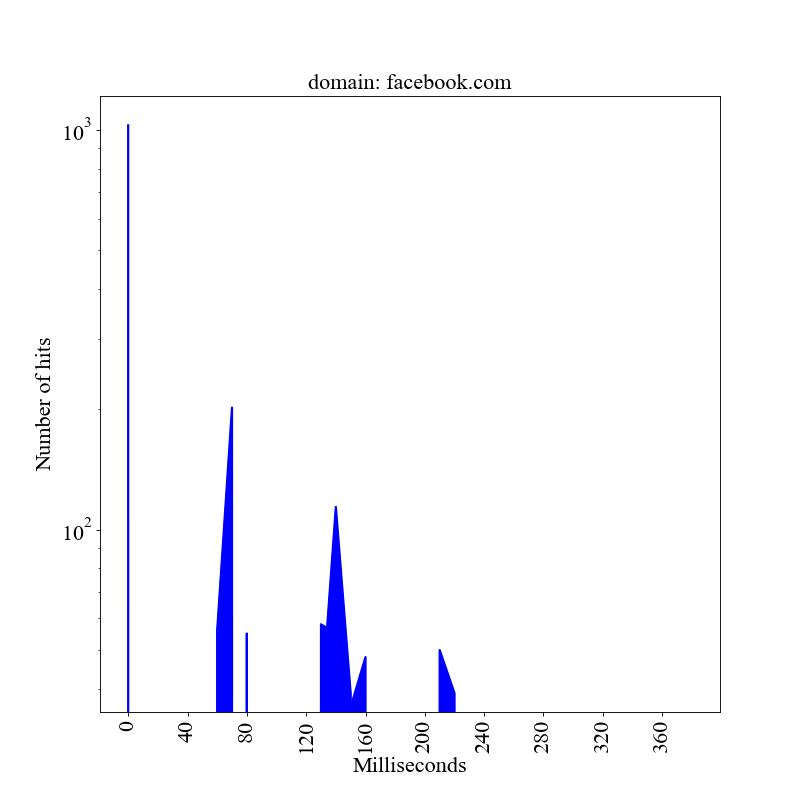}}
    \subfloat[Baidu.com distribution]{\includegraphics[width=2.5in]{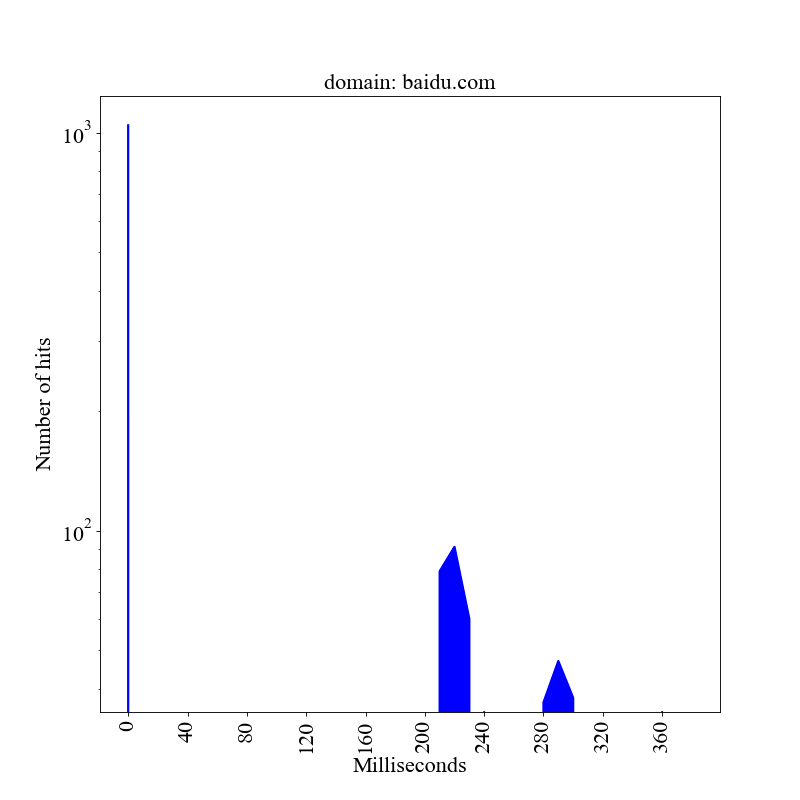}}\\ 
    \subfloat[Youtube.com resolve levels]{\includegraphics[width=2.5in]{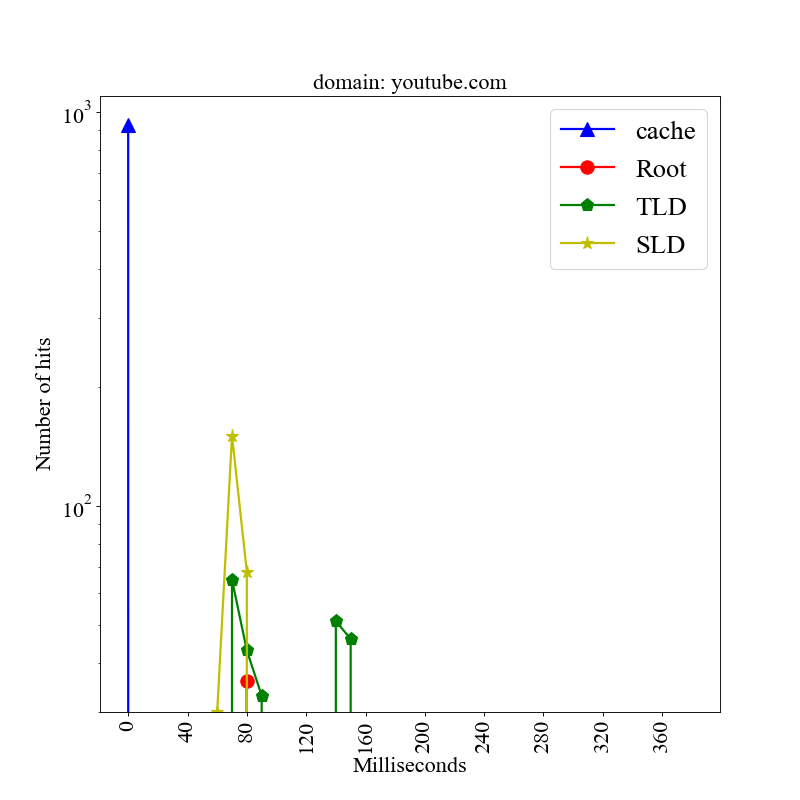}}
    \subfloat[Facebook.com resolve levels]{\includegraphics[width=2.5in]{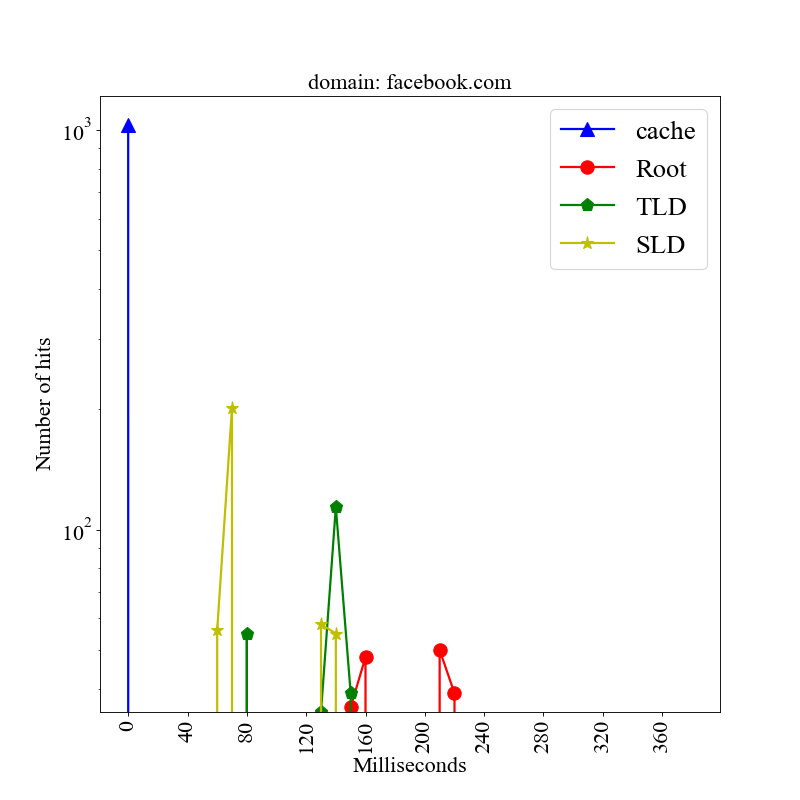}}
    \subfloat[Baidu.com resolve levels]{\includegraphics[width=2.5in]{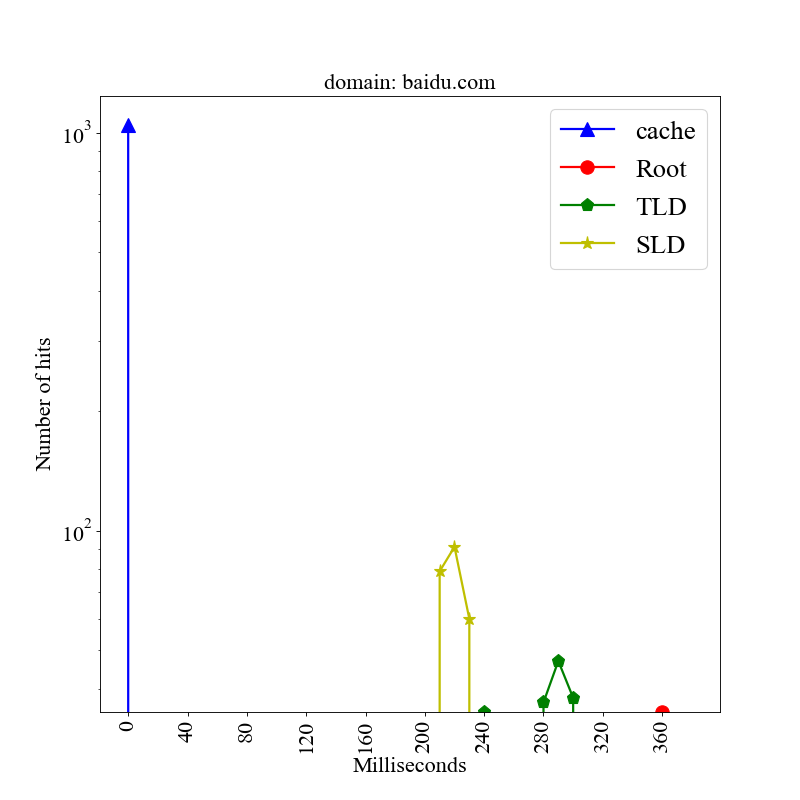}}
    
    \caption{Distribution of RTT values for the top 3 sites, in the local experiment. Each histogram is depicted at 10 ms intervals, between 0-400 ms. We addressed bins with fewer than 30 responses (out of a total of $\sim$ 1500 packets for each domain) as noise. Sub figures (a-c) support the assumption (see Fig. \ref{fig:assume}) that a proper description of the RTT value distribution is a collection of distributions with gaps between them. Sub figures (d-f) depict each DNS resolve level. They confirms the hypothesis of a correlation between the Poissons and the resolve levels. }
	\label{fig_exams}
\end{figure*}
The value measured at each level was also taken into account at the next level as a result of the recursive process. For example, a TLD's RTT value was assigned for its resolution query, and was summed with the root RTT value to create the accumulated RTT value of the root level. The process time between the query for the TLD and the root was summed as well, to obtain the actual RTT value.
Each domain was inspected in two ways:
\begin{enumerate}
	\item Level: Each DNS level was clustered in a different set.
	\item Probabilistic: This inspection mapped the common use of the recursive resolver. This inspection was divided into two parts: domain and interval.
	\begin{enumerate}
		\item \textbf{Domain:} For each domain, the data were grouped by DNS level, to obtain a percentile division. For example for the domain
		quora.com, we obtained a cache of 98.8713\%, a TLD of 0.2257\%, and a SLD of 0.9029\%.
		\item \textbf{Interval:} This inspection was specific to a time interval. For example, for the time interval of 60-70 ms of abc.com, there were 100 responses out of a total of 50,000 for the domain, with the following distribution:
		Root - 10 responses, TLD - 40 responses, SLD - 100 responses.
		In other words, the probability of obtaining responses in this interval was 0.3\%. The following probabilities corresponded to the number of responses by each level in this time interval:
		Root - 6.66\%, TLD - 26.66\%, SLD - 66\%.
	\end{enumerate}
\end{enumerate}

The RTT distribution of 3 of the top sites is depicted as an example in Fig. \ref{fig_exams}(a-c). The figures represent sites from the local experiment between 0-400 ms, at 10 ms intervals. The histograms resemble Poissons with regard to each site separately. Fig. \ref{fig_exams}(d-f) depicts the same sites with identification of each resolve level. Each Poisson is dominated by a different color, which means a different resolve level. This correlation confirms the hypothesis with regard to each domain. The gaps that can be seen between the Poissons are the baseline for our attack identification method.
\section{Identification of an attack}
\label{Ident}
As stated above, we used our methodology to generate distributions of the data. These distributions were then used to identify DNS poisoning attacks. In following section, we discuss the identification method which is identical for both DNS poisoning and DNS cache poisoning except for the place of the attacker and the defender. In our model associated with DNS poisoning, both the attacker and defender reside in the LAN. In cache poisoning attack, the attacker is placed somewhere in the internet. The defender is located beside the resolver.

In Sections \ref{Ident} and \ref{Learning}, we used the domain Yahoo.com as an example since Yahoo.com is one of the top 5 most searched domains in alexa.com. Based on the distribution of our data, we tried to evaluate our hypothesis on attack identification.
We used a heuristic function to obtain clear insights into the distribution of the data. Then, we used a DNS attack tool to generate attack packets. Notations appear in Table \ref{table:symbol}. 
\begin{table}[!]
	\caption{List of symbols} 
	\label{sym}
	\begin{tabular}{p{0.5cm}p{7cm}} 
		\hline 
		$N$ & Total number of packets (for the domain).  \\
		$t$ & Specific time interval t.  \\
		$n_t$ & Number of responses from the resolver in interval t.  \\
		$\alpha$ & Amount of data in a specific bin out of the entire data.
		\\
		$H(n)$ & Heuristic function implied on n.\\
		$B$ & Binary vector describing the impact of $H(n)$ on our data bins.\\
		$B_i$ & I'th bit of $B$.
	\end{tabular}
	\label{table:symbol} 
\end{table}
\subsection{Measurements}
\label{measure}
To assess the effectiveness of the attack, we assumed that whenever a query is sent to the recursive resolver, a single attack response is sent in parallel to the response from the recursor. The attacker's goal is to answer any query it sees as fast as it can. Therefore, it produces single packet for a query and proceeds to the next query. There were no duplicate responses or response failures. Each response had a race condition: attack response vs. the recursive resolver's response.  We describe a number of measurements, and then present our theoretical method to test our attack mitigation rate

.

\begin{figure}[!t]
	\centering
	\includegraphics[width=1.01\columnwidth]{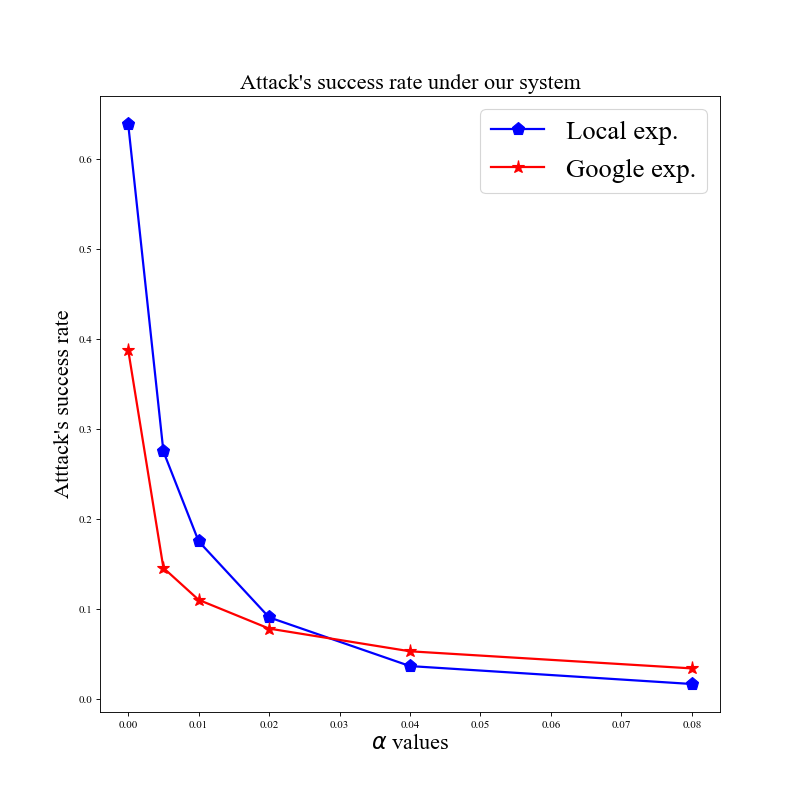}
	\caption{Success rate of an attack as a function in alpha values in our data of two experiments (which are fully described in Section \ref{Expo}). We used alpha values between 0-8\%. Each alpha value produced a different attack success rate. The identification is based on the bins where there were data exceeding the current alpha value. For example, $\alpha=0.5$ produced a $~29\%$ success rate for the attack in the Local exp., and $~19\%$ for the Google exp.}
	\label{fig:Alphas}
\end{figure}

\begin{enumerate}
	\item Probability of RTT value:  $P_r(rtt=t)$ represents the probability of getting a specific RTT value $t$. We denote by $n_t$ the number of packets in a specific interval. $N$ represents the total number of packets for the specific domain. We used the formula:
    \begin{equation} \label{eq6}
\begin{split}
P_r(rtt=t) & =\frac{n_t}{N} \\
\end{split}
\end{equation} An example of this probability distribution is presented in Fig. \ref{fig:prob}. It depicts the probability to getting an RTT value in intervals of 10 ms. It can be seen that the probability of obtaining a response in 0-10 ms. is above 50\%. The probability of obtaining a response between 60-70 ms. is $~8\%$. Each domain was described by this kind of histogram.
\begin{figure}[!h]
	\centering
	\includegraphics[width=1.01\columnwidth]{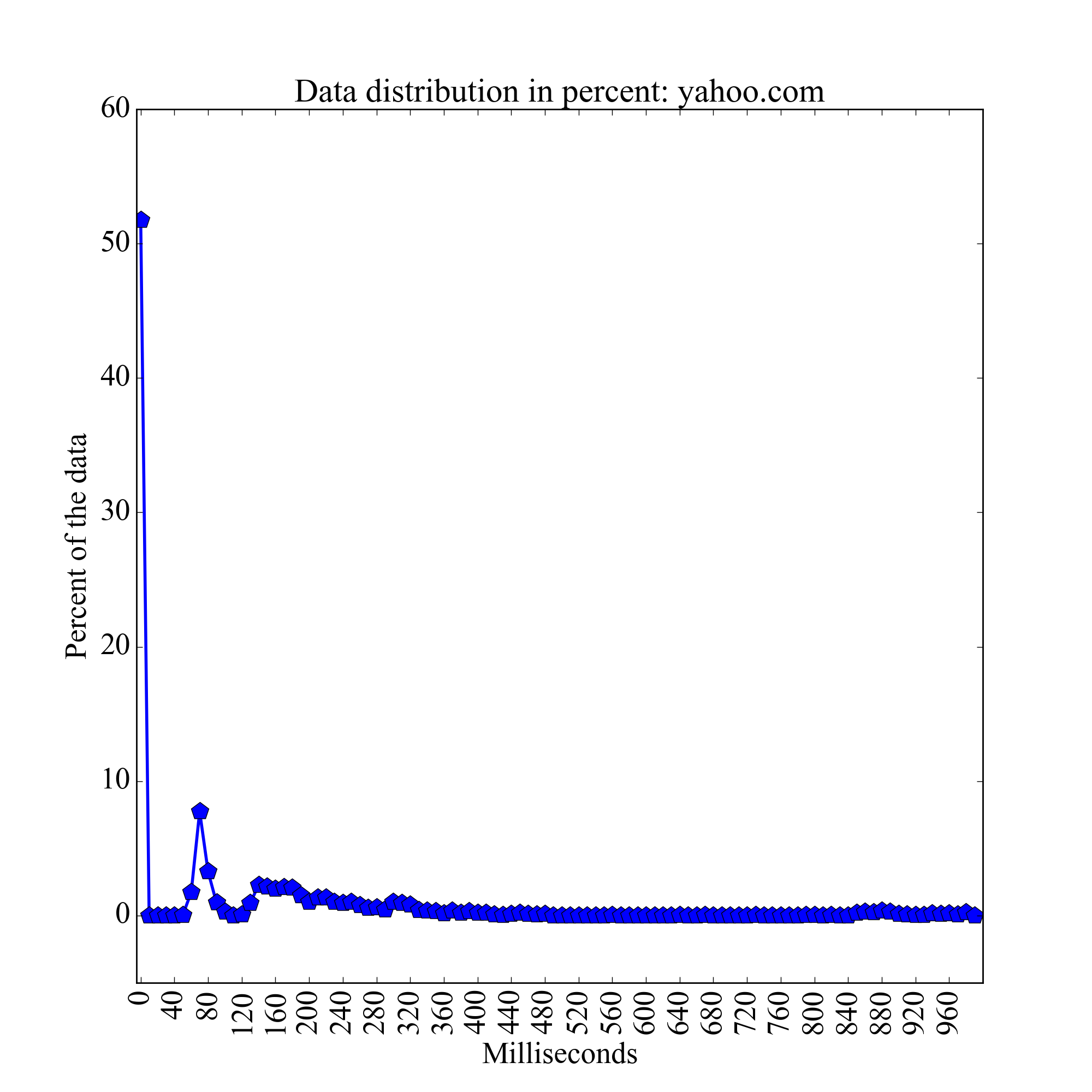}
	\caption{An example of a probability distribution: Yahoo.com which is one of the top 5 domains from alexa.com, at 10 ms intervals. More than 50\% of the values are from the cache in the first bin between 0-10 ms. The remaining values are centered around 70-90 ms, and 130-200 ms.}
	\label{fig:prob}
\end{figure}
	\item Probability of attack success: As mentioned in Section \ref{model}, we used an eavesdropper attack that can inject counterfeit packets. To simplify our calculations, we assumed that the attack can fake response to any query it sees.  Therefore, the attack success' probability is 100\%. Thus, a change in the number of packets the attack sends does not alter its success rate.

\item Heuristic function:
\label{alpha}
To document the success rate of the attack with and without our system, we took the percentile distribution from Section \ref{measure}(1).
Our purpose was to remove negligible amounts of RTT values to obtain a clearer picture of each domain. Thus, we mapped each bin with the function $H(n)$, where $\alpha$ is the percentage of the data in the specific bin.

\[H(n) = 
\begin{cases*}
1, & if $ \alpha > threshold $ \\
0, &  elsewhere.  \\
\end{cases*}\]

\item New attack success rate: Applying $H(n)$ on our data produced a 1D binary vector. We denote it as $B$. Its length is 100 since every bin describes 10 ms. for an interval of 0-1000 ms. This is the interval we got responses in. We summed all the intervals, and averaged the success on the complete interval by dividing the sum by 100. Therefore, the attack success rate with our system is:

\begin{equation} \label{eq3}
\begin{split}
P_r[attack's\; success] =0.01*\sum_{i=1}^{100}H(n)*B_i
\end{split}
\end{equation}

\end{enumerate}
We tried multiple alpha values between 0\%-8\%. Fig. \ref{fig:Alphas} depicts the influence of the alpha values on the attack's success rate in our system. A higher alpha value results in a higher mitigation rate/lower attack's success rate. However, a higher alpha value increases the false positive (FP) rate  where many benign packets are falsely indicated as attacks. An extreme case is $\alpha=100\%$ which erases all the data, and marks every packet as an attack. In this case, the false negative (FN) rate is 0\% but the FP is 100\%. For a $\alpha$ value exceeding 0.5 the attack's success rate drops by 70\%-80\%.

\hl{}

\subsection{Experimental attack}
\label{Exp_at}
We also tested an experimental attack. This attack was created by our DNS poisoning tool\cite{attacker_tool}. It was executed as an inside-the-LAN attack between the client and the recursive resolver. 
The distribution of the attack was between 0-20 ms, since we designed it to outrun the recursor and its cached memory responses. Because there was a considerable difference between the RTT values of the attack responses and the responses from the resolution process, we only used the responses tagged as cache as benign data for our identification method.

\begin{figure}[!]
	\centering
	\includegraphics[width=1.01\columnwidth]{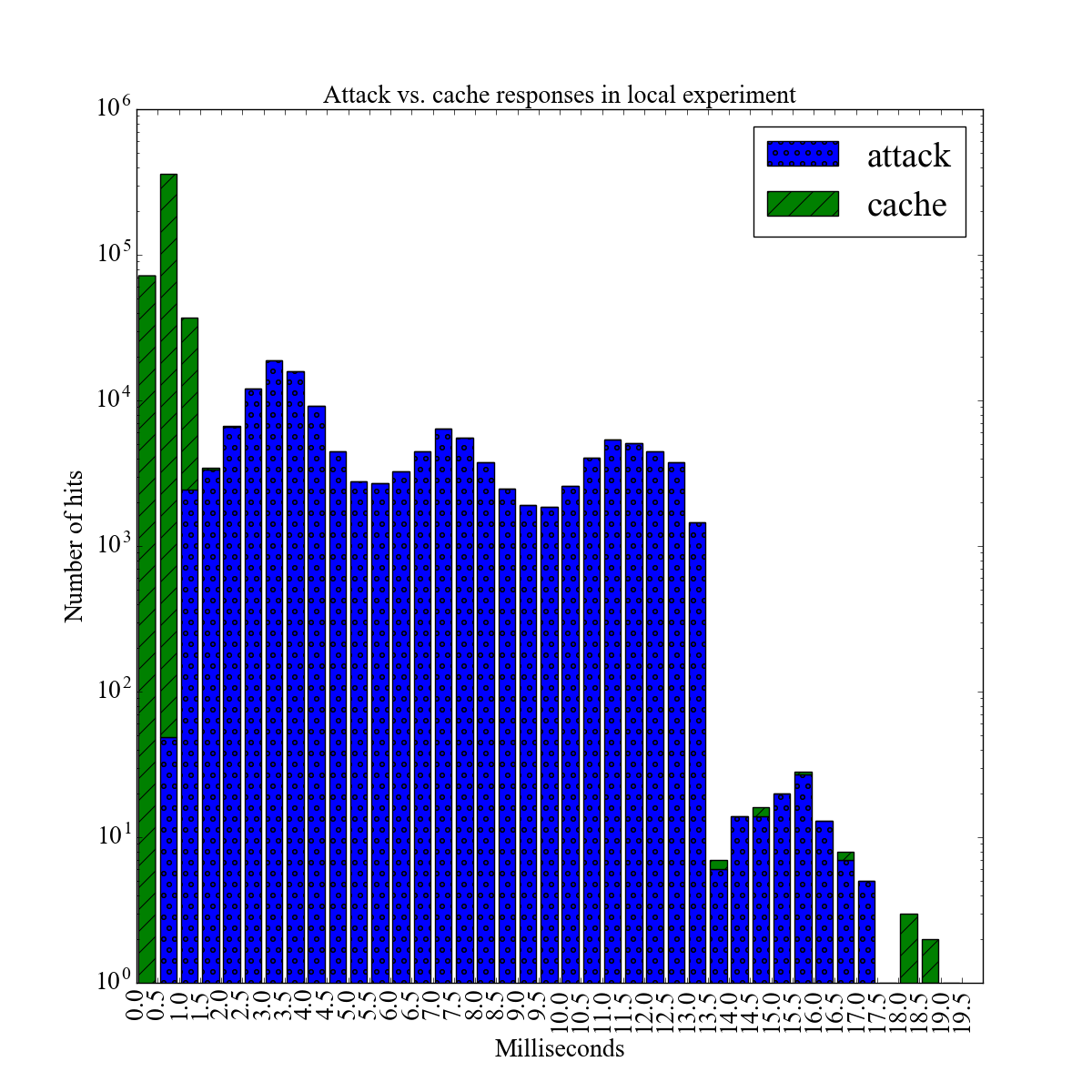}
	\caption{Attack and cache distribution from the local environment, between 0-20 ms at 0.5 ms intervals. The intersection of the attack and the cache responses is in the second and third bins. Thus tagging these bins as cache produces a 0.66\% error rate.}
	\label{fig:at_c_lo}
\end{figure}

\begin{figure}[!]
	\centering
	\includegraphics[width=1.01\columnwidth]{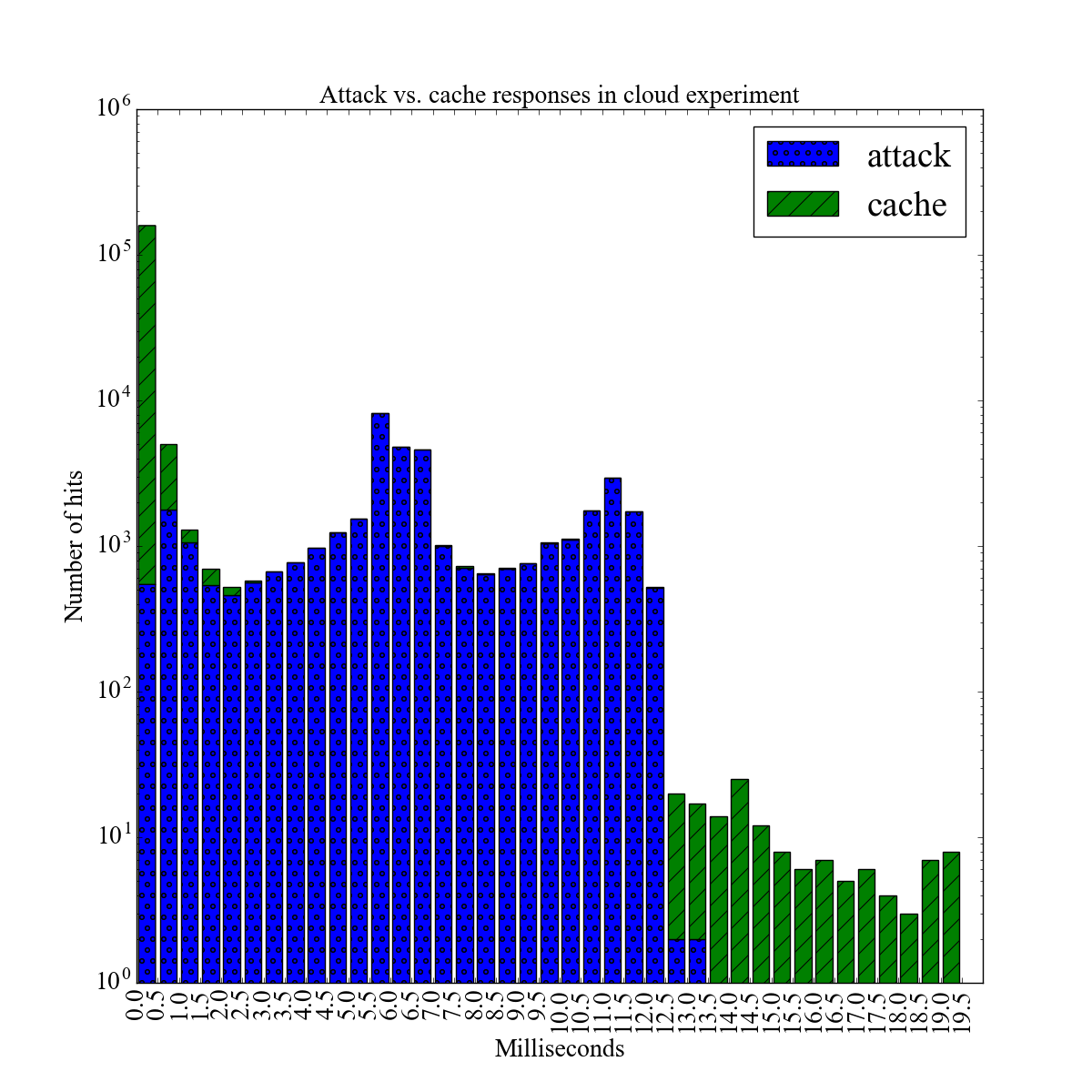}
	\caption{Attack and cache distribution from the cloud environment, between 0-20 ms at 0.5 ms intervals. The intersection of the attack and the cache responses is in the first and second bins. Thus tagging these bins as cache produces a 1.4\% error rate.}
	\label{fig:at_c_go}
\end{figure}

We first tested the intuitive notion of a threshold for the cache by generating histograms of the data from our experiment(Fig. \ref{fig:at_c_lo} for local environment  and Fig. \ref{fig:at_c_go} for cloud environment).

Intuitively the bins where the cache and attack responses intersect, and where the cache constitutes the majority of each bin should be tagged as cache. The intersection of the cache and the attack responses were in the second and third bins in the local environment, and in the first and second bins in the cloud environment. In the local environment, the intuitive process yielded a 0.66\% error rate. In the cloud, it yielded a 1.4\% error rate. The results are presented in section \ref{exp_at}.

\section{Machine learning}
\label{Learning}
Our methodology and attack identification in Sections \ref{Meth} and \ref{Ident} consisted of classifying packets into two or more classes. This made it possible to implement learning machines. A successful classifying rate between intervals in any domain depends on the information acquired about that specific domain. 

In an interval where there were high values of diverse DNS levels, the classifying rate was lower. In an interval where there were a high number of benign responses, the identification rate of the attack was lower.
The gaps with no responses at all corresponded to places where the identification rate of the attack was high. 

Supervised learning\cite{kotsiantis2007supervised}, \cite{gentleman2008supervised} involves creating a function from labeled training data. This is done by using training data that consists of a set of training examples. Each train example pairs an input object and an output value. A supervised learning algorithm analyzes the training data and produces a function which can be used for mapping new samples into corresponding output values.

\textbf{Random Forest}\cite{ho1995random,ho1998random} consists of an ensemble of k untrained Decision Trees (trees with only a root node). The following steps are carried out on these roots:
at the current node, randomly select p features from available features. The number of features p is usually much smaller than the total number of features.
Then compute the best split point for tree k and split the current node into two sons. Reduce the number of total features from this node on. Repeat the previous steps until either a maximum tree of depth l has been reached or the splitting metric reaches some extremum. Repeat all the above for each tree k in the forest. Aggregate on the output of each tree in the forest.

\textbf{KNN classification algorithm}\cite{cover1967nearest,lihua2006study,dasarathy1991nearest} is used to classify a test sample according to its K training
samples that are the nearest to it. Each train sample has a label. The algorithm takes the label of the majority of the K nearest samples and assigns it to the test sample.

We chose these methods since random forest may find a set of simple rules that will accommodate the data distribution and K nearest neighbors is an algorithm that finds outliers from a main distribution. It turned out to be more stable than other anomaly detection algorithms we tried.

We used these learning methods to obtain the best identification rate. As a ground rule for these learning machines, we used a ratio of 80/20 between the training and test samples.
\section{Experiments}
\label{Expo}
In this section, we present the technical specifications of the data collection for Sections \ref{Meth} and \ref{Ident}. We ran experiments in two environments: local and in the cloud. The cloud provider we used was Google's cloud engine. We used local machines for the lab experiment and VMs for the cloud experiment. We used a network bandwidth of 100Mb. Both experiments were run in $\sim$ 4-7 days. We took the top 500 sites from alexa.com as a sample to simulate user access to different sites.

The software we used to imitate a DNS recursive resolver was BIND\cite{bind9}. We used tshark\cite{tshark} software to log the communication both in the client and the resolver. We analyzed the domain, RTT value, query/response flag, ip src/dst, and the answer itself. The identifying process was based on an IANA TLD database\cite{iana_db}. For specific level inspection, we also used a DNS tool for Python called dnspython\cite{dnspython}, and nslookup software\cite{boyle2013applied}, by querying the dns tool and performing a nslookup query if the former failed. We used a dns-cache-poisoning tool\cite{attacker_tool} as the attack software. This is a simple open source tool which we customized easily to our experiment. We added a randomized sleep to the attack to randomize its RTT values.
The attack was detected by a Python sickit implementation of machine learning\cite{sickit}. Our data are available in a Google drive\cite{drive}. 
\begin{figure}[htp]
	\subfloat[Exp. 1]{%
  \includegraphics[clip,width=1\columnwidth]{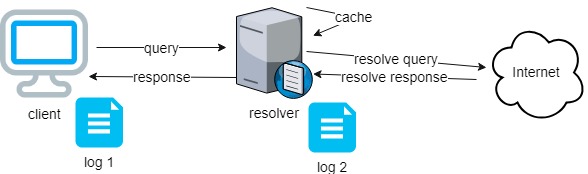}%
}

	\subfloat[Exp. 2]{%
  \includegraphics[clip,width=1\columnwidth]{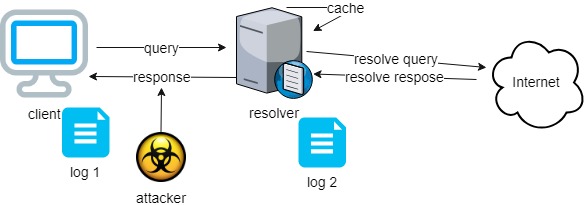}%
}
	\caption{In Exp. 1 a client queries a recursive resolver, and the communication is recorded in log files(log 1 and log 2). In Exp. 2 the attacker spoofs responses to the client.}
    \label{fig:client}
\end{figure}

The experiments are depicted in Fig. \ref{fig:client}; experiment 1, is shown in Fig. \ref{fig:client}(a). The experiment involved recording benign DNS packets between a stub resolver and a recursive resolver. Each query or response within the LAN between the client and the recursor was recorded by a log file(log 1). Each resolve query/response that was taken by the recursive resolver outside the LAN was recorded by another log file(log 2). We correlated the two log files by domain+ID parameters to determine the DNS level from which client received a response. 

This way, the stub resolver's RTT value was mapped onto a DNS level produced outside of the LAN/recursive resolver's log file. We needed to produce the recursor's responses from the cache and the resolution process. This was made possible by changing the TTL value in the recursive resolver. To get cache responses, we changed the TTL value to $\sim$ 3 hours. To collect more resolution responses, we changed the TTL value to 0. This way, we mapped the behavior of the recursive resolver in both cases.

Experiment 2, depicted in Fig. \ref{fig:client}(b), involved an attacker between the client and the resolver. 
Each experiment produced about $\sim 0.5 m$ response packets. 


We obtained real data from IUCC\cite{Inter}. These data were acquired from a line between Tel Aviv and Frankfurt and were recorded over about $\sim$2 days. Congestion on this link caused packet drop. In addition, uneven routing caused some of the responses to be delivered in another link. Thus, some of the queries were unanswered. Overall, we only used about $\sim$25\% of the packets, or approx. 1.3 million. We assumed that these data did not contain attempts at cache poisoning. As stated in Section \ref{Meth}, we only analyzed the IUCC data in Section \ref{Spec}. Therefore, we analyzed the IP sources and not domains from each packet.

To control for data reliability, we split the data from each data source into a ratio of 80/20 of train/test. We assessed whether the test data distributed in a similar way as the train data. We found that the distributions were similar. In the IUCC data there was less than a 5\% differentiation for each time interval between the train and test data. In the data from our experiments, most of the data had less than a 5\% differentiation in each time interval between the train and test data. Less than 10\% of the domains had 1-3 intervals which had $\sim$5\%-10\%  differentiation between the train and test data.
\section{Results}
\label{results}
In the following section, we describe the results of Sections \ref{Meth} and \ref{Ident} from the learning machines described in Section \ref{Learning}. A list of abbreviations is presented in Table \ref{abb}.

\begin{table}[!h]
	\caption{List of Abbreviations for result tables} 
	\label{abb}
	\centering 
	\begin{tabular}{c c} 
		\hline 
		Acc & Accuracy  \\
		Dev & Deviation  \\
		Var & Variance  \\
		RF & Random Forest  \\
		KNN & K Nearest Neighbors  \\
		Att & Attack's identification rate  \\
		FP & False Positives \\
		FN & False Negatives \\
		L & Local experiment \\
		C & Cloud experiment \\
		sub & subdomain \\
	\end{tabular}
	\label{table:nonlin} 
\end{table}

\subsection{Methodology}
\label{meth_res}
As presented in Section \ref{Meth}, we separated the data into specific domains, creating $\sim500$ sub-datasets. We added the deviation and variance for each indicator. We used Random Forest (RF) and K Nearest Neighbors (KNN) to identify the origin of the responses, as stated in Section \ref{Learning}. The average proportion of the cache/DNS level of resolve in each data is presented in Table \ref{Pro}.

\begin{table}[!h]
	\centering
	\caption{Proportion of packets in the data}
	\label{Pro}
	\begin{tabular}{|l|l|l|l|l|l|l|l|}
		\hline
		& Cache & Root & gTLD  &ccTLD &   sub & host               \\ \hline
		L&41.651 & 23.257 & 13.432  &1.644 &   19.87 & 0.13               \\ \hline
		C&16.76 &32.88 & 19.38  &2.872 &   28.072& 0.07               \\ \hline
	\end{tabular}
\end{table}

Tables \ref{Go1} and \ref{Lo1} present the cache and non-cache identification. This calculation was straightforward as was shown in Section \ref{Cache_or_non} since the RTT values were quite different between the cache and the resolve levels. We achieved an accuracy rate of 98\%. This was done to better identify the distribution of the DNS packets, and to test our ability to distinguish between the DNS levels.

\begin{table}[!]
	\centering
	\caption{Cache/no cache differentiation in the cloud environment.}
	\label{Go1}
	\begin{tabular}{|l|l|l|l|}
		\hline
		& Acc & Dev & Var                    \\ \hline
		RF       & 0.983                            & 0.02     & 5*$10^{-4}$                       \\ \hline
		KNN & 0.985                            & 0.02     & 4*$10^{-4}$ \\\hline
	\end{tabular}
\end{table}

\begin{table}[!]
	\centering
	\caption{Cache/no cache differentiation in the local environment.}
	\label{Lo1}
	\begin{tabular}{|l|l|l|l|}
		\hline
		& Acc & Dev & Var                    \\ \hline
		RF       & 0.983                           & 0.02    & 5*$10^{-4}$                       \\ \hline
		KNN & 0.985                            & 0.02     & 5*$10^{-4}$ \\\hline
	\end{tabular}
\end{table}

Tables \ref{Go2} and \ref{Lo2} indicate the identification rates of the DNS hierarchy levels, as stated in Section \ref{Spec}. We analyzed the DNS level and the cache data. We obtained a correct identification rate of approximately 75\% in the cloud, and $\sim$84\% in the local data. This analysis proved to be more difficult, because each domain depicts a slightly different histogram. Some created a dense distribution, with a number of different DNS levels in the center which made it hard to differentiate between levels. Furthermore, due to the cloud's resources, the RTT value was small in most of the responses. Therefore, dissecting levels in its data was less successful. 

\begin{table}[!]
	\centering
	\caption{Identifying DNS levels including cache in the cloud environment.}
	\label{Go2}
	\begin{tabular}{|l|l|l|l|}
		\hline
		& Acc & Dev & Var                    \\ \hline
		RF       & 0.75                            & 0.1     & 0.01                       \\ \hline
		KNN & 0.69                           & 0.1     & 0.01 \\\hline
	\end{tabular}
\end{table}

\begin{table}[!]
	\centering
	\caption{Identifying DNS levels including cache in the local environment.}
	\label{Lo2}
	\begin{tabular}{|l|l|l|l|}
		\hline
		& Acc & Dev & Var                    \\ \hline
		RF       & 0.81                            & 0.06     & 0.004                       \\ \hline
		KNN & 0.84                            & 0.06     & 0.003 \\\hline
	\end{tabular}
\end{table}

Tables \ref{Go4} and \ref{Lo4} show the identification rate of the DNS hierarchy levels, from the root level excluding the cache data. We obtained a correct identification rate in the cloud of about 70\%, and $\sim$76\% in the local data. Tables \ref{Go2} and \ref{Lo2} include the cache data, which are more distinguishable. Therefore, the identification rate is different in this case.

\begin{table}[!]
	\centering
	\caption{Identifying DNS levels without cache in the cloud environment.}
	\label{Go4}
	\begin{tabular}{|l|l|l|l|}
		\hline
		& Acc & Dev & Var                    \\ \hline
		RF       & 0.64                            & 0.1     & 0.01                       \\ \hline
		KNN & 0.7                           & 0.1    & 0.01 \\\hline
	\end{tabular}
\end{table}

\begin{table}[!]
	\centering
	\caption{Identifying DNS levels without cache in the local environment.}
	\label{Lo4}
	\begin{tabular}{|l|l|l|l|}
		\hline
		& Acc & Dev & Var                    \\ \hline
		RF       & 0.71                            & 0.08     & 0.007                       \\ \hline
		KNN & 0.76                            & 0.08     & 0.007 \\\hline
	\end{tabular}
\end{table}

As can be seen in the tables, the identification rate in almost every stage was slightly different between the environments. This can be attributed to disparities in Google's resources. Each packet goes through a smaller number of hops in the resolution process in the cloud than in the local resolver. Google has considerably shorter routes to most of the root/gTLD/ccTLD servers, which gives it a low RTT value for most requests. Thus, it is more difficult to separate the levels in the data from Google.

\subsection{Empirical attack}
\label{exp_at}
We ran an experimental attack, as mentioned in Section \ref{Exp_at}. The attack packets were received between 0-20 ms, so that they arrived before the cache responses. An eavesdrop/inject attack between the stub resolver and the recursive resolver was used to generate these attack packets. Since the attack was located inside the LAN, as indicated in Section \ref{model}, the attack packets arrived fast. We used its data along with the data from section \ref{meth_res}. The results are presented in Tables \ref{Go_anom} and \ref{Lo_anom}.

\begin{table}[th]
	\centering
	\caption{Experimental attack identification rate from the cloud environment. The deviation in all the cases was below 4\%.  }
	\label{Go_anom}
	
	\begin{tabular}{|>{\centering}m{1cm}|>{\centering}m{1cm}|>{\centering}m{1cm}|>{\centering}m{1cm}|>{\centering}m{1cm}|}
		\hline
		& Acc        & FP    & FN     \tabularnewline \hline
		RF       & 0.987  & 0.005 & 0.006  \tabularnewline \hline
		KNN      & 0.990  & 0.003 & 0.006  \tabularnewline \hline
	\end{tabular}

\end{table}

\begin{table}[th]
	\centering
	\caption{Experimental attack identification rate from the local environment. The deviation in all the cases was below 1\%.  }
	\label{Lo_anom}
	
	\begin{tabular}{|>{\centering}m{1cm}|>{\centering}m{1cm}|>{\centering}m{1cm}|>{\centering}m{1cm}|>{\centering}m{1cm}|}
		\hline
		& Acc      & FP     & FN     \tabularnewline \hline
		RF           & 0.997   & 0.001  & 0.001  \tabularnewline \hline
		KNN          & 0.998  & 0.001  & 0.0006 \tabularnewline \hline
	\end{tabular}

\end{table}
As can be seen in Tables \ref{Go_anom} and \ref{Lo_anom}, we had $\sim99\%$ correct identification rate. As mentioned in Section \ref{Expo}, we generated $\sim0.5 m$ attack packets. To be precise, we identified 355,606  out of 359,197 attack packets from the Google data, and 131,415 out of 131,810 from our local simulation data. We assumed that although the identification rate for the empirical attack was superior to the naive calculation, the difference was not pronounced since each learning machine had only one feature. Thus, the machine's power was comparable to the naive threshold. Our identification rate was found to be superior to the naive threshold by 0.2\%-0.4\%. 
\section{Conclusion}
\label{conclu}
In this paper we presented an innovative method to detect DNS poisoning attacks. We assumed that the behavior of the RTT value could be generalized as a number of Poisson distributions. Each Poisson addresses a resolve level. Thus, analyzing the gaps between the Poissons can serve to detect attacks. We confirmed our hypothesis in various environments, and used it as a basis for identification of the attack.

This study presented our method for an experimental local network attack. In the future, we aim to apply this identification method to other kinds of attacks. For example, it could be used to identify cache poisoning attacks against recursive resolvers. To do so, each level will be inspected separately to detect anomalies from the recursor's point of view. This appears to be easier, since no classifying process is needed, given the lack of intersecting levels. The time analysis may be more precise, since the distribution of only one DNS level is more distinguishable.

Future work will also concentrate on the precision of the analysis. In this paper, we analyzed the data at milliseconds intervals. We saw that this process failed to fit Google. In the future, we will attempt to analyze the data at microseconds intervals to achieve higher accuracy, while determining the level of precision that results in a fit and avoids overfitting.

Adding packet drops, multiple responses and other kinds of failures may change our model. These cases are intriguing topics for further investigations and may make the results reported here more resilient. These models may lead to a prototype for a more realistic defense system. 

\end{document}